\documentstyle[prb,aps,floats,epsf]{revtex}
\begin{document}
\twocolumn
\title{Propagation of solitons of the magnetization
in magnetic nano-particle arrays}
\author{Satoshi Ishizaka and Kazuo Nakamura}
\address{NEC Fundamental Research Laboratories, \\
34 Miyukigaoka, Tsukuba, Ibaraki, 305, Japan}
\date{\today}
\maketitle
%%%%%%%%%%%%%%%%%%%%%%%%%%%%%%%%%%%%%%%%%%%%%%%%%%%%%%%%%%%%%%%%%%%%%%%%%%%%%%%
%
%
%
%
%
%%%%%%%%%%%%%%%%%%%%%%%%%%%%%%%%%%%%%%%%%%%%%%%%%%%%%%%%%%%%%%%%%%%%%%%%%%%%%%%
\begin{abstract}
It is clarified for the first time that solitons originating from the
dipolar interaction in ferromagnetic nano-particle arrays are stably
created.
The characteristics can be well controlled by the strength of the
dipolar interaction between particles and the shape anisotropy of the particle.
The soliton can propagate from a particle to a
neighbor particle at a clock frequency even faster than 100 GHz using materials
with a large magnetization.
Such arrays of nano-particles might be feasible in an
application as a signal transmission line.
\end{abstract}
\narrowtext
%%%%%%%%%%%%%%%%%%%%%%%%%%%%%%%%%%%%%%%%%%%%%%%%%%%%%%%%%%%%%%%%%%%%%%%%%%%%%%%
%
%
%
%
%
%%%%%%%%%%%%%%%%%%%%%%%%%%%%%%%%%%%%%%%%%%%%%%%%%%%%%%%%%%%%%%%%%%%%%%%%%%%%%%%
The recently developed nanofabrication techniques make it possible to
fabricate ferromagnetic particles to a length of 20-30 nanometers.
\cite{Honda99a}
Since the size of such nano-particles is small enough to be a single
domain, the magnetization is homogeneous over a particle and 
can be described by a magnetic moment.
These particles interact with each other due to a dipolar interaction of the
magnetic moments.
In such a nano-particle, the magnetic anisotropy energy can be
dominated by the particle's shape rather than the magnetocrystalline
anisotropy.
Therefore, the magnetic anisotropy energy can be well controlled by changing
the shape of the particle.
When the distance between particles is short enough that dipolar interaction
becomes comparable with the anisotropy energy, 
the direction of the magnetization of each particle is determined in order
to minimize both energies, and 
the direction of the magnetization moves in a corrective manner among the
particles.
\par
%%%%%%%%%%%%%%%%%%%%%%%%%%%%%%%%%%%%%%%%%%%%%%%%%%%%%%%%%%%%%%%%%%%%%%%%%%%%%%%
%
%%%%%%%%%%%%%%%%%%%%%%%%%%%%%%%%%%%%%%%%%%%%%%%%%%%%%%%%%%%%%%%%%%%%%%%%%%%%%%%
It is known that the magnetic domain wall in
bulk magnetic materials is a kind of solitons of nonlinear waves.
\cite{Enz64a,Mikeska77a,SolitonA}
Usually, the soliton originates from the exchange interaction between spins.
This letter clarified for the first time that,
in a one-dimensional array of nano-particles,
solitons originating from the dipolar interaction between particles
are present, and that
the characteristics can be well controlled by the strength of the
dipolar interaction and the shape anisotropy.
The great advantage of control is quite a contrast to the
case of the soliton of the domain walls in bulk materials.
\par
%%%%%%%%%%%%%%%%%%%%%%%%%%%%%%%%%%%%%%%%%%%%%%%%%%%%%%%%%%%%%%%%%%%%%%%%%%%%%%%
%
%%%%%%%%%%%%%%%%%%%%%%%%%%%%%%%%%%%%%%%%%%%%%%%%%%%%%%%%%%%%%%%%%%%%%%%%%%%%%%%
In this letter, we analyze the characteristics of the solitons in arrays
of ferromagnetic nano-particles by varying the experimentally controllable
parameter: the shape anisotropy energy.
When the damping is small, the soliton propagates from a particle to a
neighbor particle at 50 GHz, even faster than 100 GHz using materials with
a large magnetization.
Such arrays of nano-particles might be feasible in an application as a signal
transmission line.
\par
%%%%%%%%%%%%%%%%%%%%%%%%%%%%%%%%%%%%%%%%%%%%%%%%%%%%%%%%%%%%%%%%%%%%%%%%%%%%%%%
%
%%%%%%%%%%%%%%%%%%%%%%%%%%%%%%%%%%%%%%%%%%%%%%%%%%%%%%%%%%%%%%%%%%%%%%%%%%%%%%%
We examine a one-dimensional array of particles in the $x$-direction.
Each particle has a magnetic moment $\vec M_i$ ($|\vec M_i|\!=\!M$).
The dipolar interaction and anisotropy energy is modeled as
\begin{eqnarray}
H&=&\sum_{\langle ij \rangle}
\frac{\mu_0}{4\pi r_{ij}^3}
[ \vec M_i \cdot \vec M_j
-\frac{3}{r_{ij}^2}
 (\vec M_i \cdot \vec r_{ij}) (\vec M_j \cdot \vec r_{ij})
] \cr
&+&\sum_i\frac{1}{M}[-K_y M_{iy}^2+K_z M_{iz}^2],
\label{eq: Hamiltonian}
\end{eqnarray}
where $K_y$ and $K_z$ describe the shape anisotropy.
The dipolar interaction is characterized by $J\!=\!\mu_0 M^2/(4\pi a^3)$.
In this letter, we consider the case of $K_z\!\gg\!J$ and $K_z\!\gg\!K_y$,
and thus the $x\!-\!y$ plane is an easy plane for the magnetization.
Further, we consider the case where the thermal fluctuation in the direction of
$\vec M_i$, which frequently leads to the super-paramagnetism,
\cite{Bean56a}
is well-suppressed by $K_y$ and/or $J$.
As shown below, the condition can be sufficiently satisfied since
$J$ can exceed $10,000$ K even for particles of a size of 20 nm,
when these are aligned closely to each other.
\par
%%%%%%%%%%%%%%%%%%%%%%%%%%%%%%%%%%%%%%%%%%%%%%%%%%%%%%%%%%%%%%%%%%%%%%%%%%%%%%%
%
%%%%%%%%%%%%%%%%%%%%%%%%%%%%%%%%%%%%%%%%%%%%%%%%%%%%%%%%%%%%%%%%%%%%%%%%%%%%%%%
The magnetization of each nano-particle obeys the Bloch equation:
\cite{Gilbert55}
\begin{equation}
\frac{d \vec M_i}{dt}= -\frac{\nu}{M}\vec M_i \times
(-\frac{\partial H}{\partial\vec M_i}
-\frac{\alpha M}{\nu}\frac{d \vec M_i}{dt}),
\label{eq: Bloch}
\end{equation}
where $\nu=g\mu_B/\hbar$ is a gyromagnetic constant, 
and $\alpha$ describes the dumping due to energy dissipation,
such as magnetoelastic dissipation
and eddy current loss in metallic ferromagnets.
In nano-particles, however, it has been shown that the amount of these
dissipation mechanisms are negligibly small. \cite{Tatara94a,Grag89a}
In fact, considering the eddy current loss, 
$\alpha\!\sim\!10^{-4}$ for Fe-particles of a size of 20 nm, for which
the soliton can propagate beyond 1,000 sites.
Therefore, we exclusively consider the case of $\alpha\!=\!0$ in this letter.
\par
\begin{figure}
\epsfxsize=7.0cm
\centerline{\epsfbox{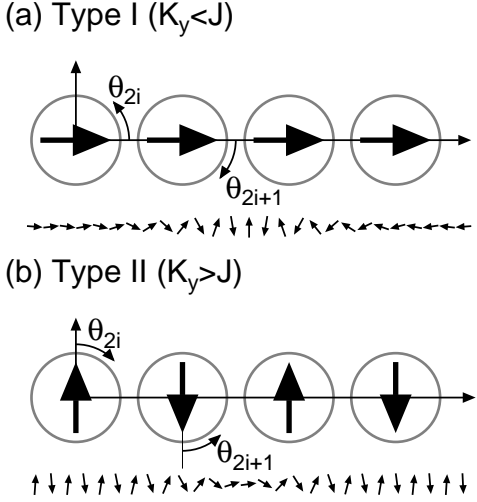}}
\caption{
The configuration of the magnetization giving a minimum total energy.
(a) type I for $K_y\!<\!J$ and (b) type II for $K_y\!>\!J$.
For both cases, an example of the soliton is shown below each figure.
}
\label{fig: Configure}
\end{figure}

%%%%%%%%%%%%%%%%%%%%%%%%%%%%%%%%%%%%%%%%%%%%%%%%%%%%%%%%%%%%%%%%%%%%%%%%%%%%%%%
%
%%%%%%%%%%%%%%%%%%%%%%%%%%%%%%%%%%%%%%%%%%%%%%%%%%%%%%%%%%%%%%%%%%%%%%%%%%%%%%%
Before discussing the results of the numerical simulations of
Eq.\ (\ref{eq: Bloch}), it is worth considering the continuum limit.
There are two types of configurations of $\vec M_i$ giving the minimum
total energy depending on $K_y$.
When only the nearest-neighbor interaction is considered
for simplicity, the configuration is 
type I [Fig.\ \ref{fig: Configure} (a)] and 
type II [Fig.\ \ref{fig: Configure} (b)] for $K_y\!<\!J$ and $K_y\!>\!J$,
respectively.
In both cases, the Bloch equation (\ref{eq: Bloch}) is approximated
to the sine-Gordon (SG) equation: \cite{SolitonB}
\begin{equation}
\nabla^2\theta'(x,t)
-\frac{1}{c^2}\frac{\partial^2\theta'(x,t)}{\partial t^2}
-\frac{1}{\lambda^2}\sin\theta'(x,t)=0,
\label{eq: Soliton}
\end{equation}
where $c^2=3Ja^2K_z\nu^2/M^2$, $\lambda^2=3Ja^2/(4|K_y-J|)$, and
$\theta'(x,t)/2=\theta(x,t)$ is the azimuthal angle of $\vec M(x)$
in the continuum limit of $\vec M_i$.
Note that the definitions of $\theta_i$ for even- and odd-sites are different
as shown in Fig.\ \ref{fig: Configure}.
Consequently, the solution in the form of the soliton is present for both
cases of type I and II, whose examples are shown below 
Figs.\ \ref{fig: Configure} (a) and (b).
The solution has a velocity ($v$) as a parameter, and its wavelength
is $l\!=\!\lambda\sqrt{1-v^2/c^2}$.
The maximum velocity is limited to $c$.
\par
%%%%%%%%%%%%%%%%%%%%%%%%%%%%%%%%%%%%%%%%%%%%%%%%%%%%%%%%%%%%%%%%%%%%%%%%%%%%%%%
%
%%%%%%%%%%%%%%%%%%%%%%%%%%%%%%%%%%%%%%%%%%%%%%%%%%%%%%%%%%%%%%%%%%%%%%%%%%%%%%%
%%%%%%%%%%%%%%%%%%%%%%%%%%%%%%%%%%%%%%%%%%%%%%%%%%%%%%%%%%%%%%%%%%%%%%%%%%%%%%%
%
%%%%%%%%%%%%%%%%%%%%%%%%%%%%%%%%%%%%%%%%%%%%%%%%%%%%%%%%%%%%%%%%%%%%%%%%%%%%%%%
When the long range part of the interaction is taken into account,
$|K_y\!-\!J|$ in $\lambda^2$ is replaced by 
$|K_y\!-\!5\zeta(3)J/4|\!\approx\!|K_y\!-\!1.5J|$
with $\zeta(x)$ being Riemann's zeta function.
To gain a large value of $J$, however, it is better to align
particles closely in such a way that the distance between particles ($a$)
is comparable with the diameter of the dot ($d$), where the nearest-neighbor
interaction becomes dominant.
Therefore, we only take into account the nearest-neighbor interaction
in Eq.\ (\ref{eq: Hamiltonian}) hereafter, although the actual form of the
interaction may slightly deviate from that in Eq.\ (\ref{eq: Hamiltonian}) for
$a\!\sim\!d$.
\par
\begin{figure}
\epsfxsize=7.5cm
\centerline{\epsfbox{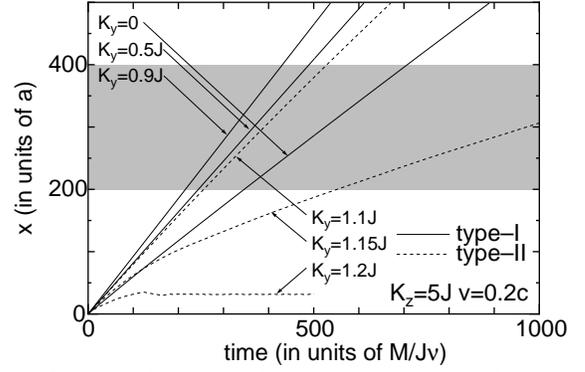}}
\caption{
The time-evolution of the center coordinate
of the soliton for various values of $K_y$.
Solid and dotted lines are for type I and type II, respectively.
A gray zone shows the range ($200\!\le\!x\!\le\!400$) where
the average velocity plotted in Fig.\ \protect\ref{fig: Velocity} (a) is
obtained.
}
\label{fig: Evolution}
\end{figure}

%%%%%%%%%%%%%%%%%%%%%%%%%%%%%%%%%%%%%%%%%%%%%%%%%%%%%%%%%%%%%%%%%%%%%%%%%%%%%%%
%
%%%%%%%%%%%%%%%%%%%%%%%%%%%%%%%%%%%%%%%%%%%%%%%%%%%%%%%%%%%%%%%%%%%%%%%%%%%%%%%
Figure \ref{fig: Evolution} shows the time-evolution of the center coordinate
of the soliton obtained from the numerical calculations of
Eq.\ (\ref{eq: Bloch}) for various values of $K_y$.
As an initial condition for $t\!=\!0$, we choose a solution of SG-equation
(\ref{eq: Soliton}) with an initial velocity $v\!=\!0.2c$.
For both cases of type I and II, the soliton stably propagates keeping
the initial velocity when $K_y$ is close to $J$.
The velocity begins to decrease when $|K_y\!-\!J|$ becomes large, and
the amount of the decrease is more significant for the case of type II than
type I.
The soliton for type II stops after the propagation beyond 30 sites for
$K_y\!=\!1.2J$, while the soliton for type I can propagate even for
$K_y\!=\!0$.
Therefore, the configuration of type I with $K_y$ close to $J$ is suitable
for the stable propagation of the soliton.
\par
%%%%%%%%%%%%%%%%%%%%%%%%%%%%%%%%%%%%%%%%%%%%%%%%%%%%%%%%%%%%%%%%%%%%%%%%%%%%%%%
%
%%%%%%%%%%%%%%%%%%%%%%%%%%%%%%%%%%%%%%%%%%%%%%%%%%%%%%%%%%%%%%%%%%%%%%%%%%%%%%%
\begin{figure}[tb]
\epsfxsize=7.5cm
\centerline{\epsfbox{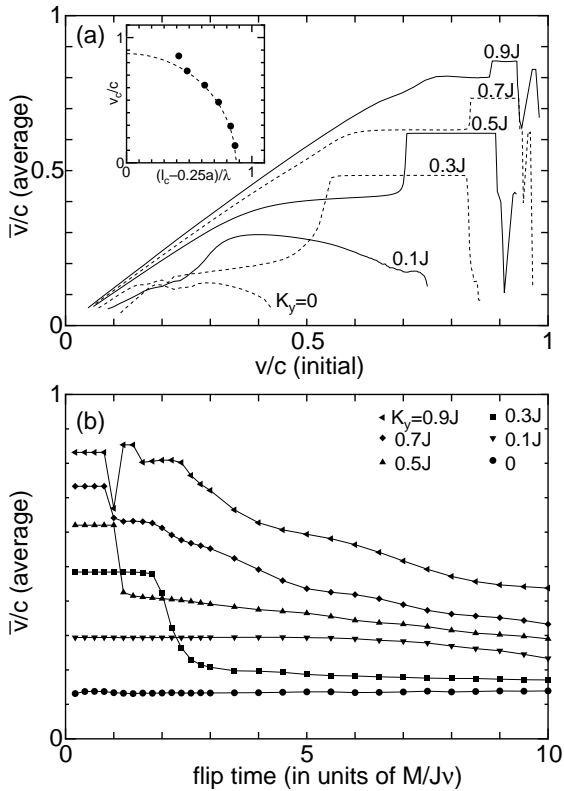}}
\caption{
(a) The average velocity obtained for the range $200\!\le\!x\!\le\!400$ as a
function of initial velocities. The inset shows the relation between the
maximum of the average velocity ($v_c$) and the corresponding wavelength
($l_c$).
(b) The average velocity of the soliton generated by flipping $\vec M_0$
as a function of the flipping time.
The average is performed in the range of $200\!\le\!x\!\le\!400$.
}
\label{fig: Velocity}
\end{figure}

%%%%%%%%%%%%%%%%%%%%%%%%%%%%%%%%%%%%%%%%%%%%%%%%%%%%%%%%%%%%%%%%%%%%%%%%%%%%%%%
%
%%%%%%%%%%%%%%%%%%%%%%%%%%%%%%%%%%%%%%%%%%%%%%%%%%%%%%%%%%%%%%%%%%%%%%%%%%%%%%%
We performed the same calculations for the type I case for various values of
the initial velocity, and obtained the average velocity ($\bar v$) in the
range, for example, of $200\!\le\!x\!\le\!400$.
The results are shown in Fig.\ \ref{fig: Velocity} (a).
The average velocity $\bar v$ keeps the initial velocity for smaller
velocities.
This implies that the solution of the continuum limit is appropriate
for this range of the velocity. 
With the increase of the initial velocity, however, the effects of the
discreteness of the array become important, and
$\bar v$ begins to saturate and comes to depend little on the initial velocity.
With the further increase of the initial velocity, $\bar v$ for
$K_y\!\ge\!0.3J$ increases suddenly with a finite jump, and $\bar v$ takes
an almost completely fixed maximum value.
It is interesting to note that the corresponding wavelength of the soliton is
always close to the distance between particles ($a$) independent of
$K_y$, while the corresponding fixed value of $\bar v$ depends on $K_y$.
In the case of $K_y\!\le\!0.1J$, although the jump becomes very broad and
the velocity is not fixed so rigorously, similar behavior also remains.
It should be noted that the sudden jump and fixing in the velocity is absent
in the case of type II.
\par
%%%%%%%%%%%%%%%%%%%%%%%%%%%%%%%%%%%%%%%%%%%%%%%%%%%%%%%%%%%%%%%%%%%%%%%%%%%%%%%
%
%%%%%%%%%%%%%%%%%%%%%%%%%%%%%%%%%%%%%%%%%%%%%%%%%%%%%%%%%%%%%%%%%%%%%%%%%%%%%%%
%%%%%%%%%%%%%%%%%%%%%%%%%%%%%%%%%%%%%%%%%%%%%%%%%%%%%%%%%%%%%%%%%%%%%%%%%%%%%%%
%
%%%%%%%%%%%%%%%%%%%%%%%%%%%%%%%%%%%%%%%%%%%%%%%%%%%%%%%%%%%%%%%%%%%%%%%%%%%%%%%
According to the numerical results, the maximum of the average velocity ($v_c$)
and the corresponding wavelength ($l_c$) satisfies an approximate relation
similar to the continuum limit:
\begin{equation}
(\frac{v_c}{c})^2+(\frac{l_c-0.25a}{\lambda})^2=0.76,
\label{eq: Dispersion}
\end{equation}
as shown in the inset of Fig.\ \ref{fig: Velocity}.
Therefore, due to the the effects of the discreteness of the array,
the wavelength is shifted by a constant, and $c$ and $\lambda$ are 
renormalized simultaneously, although the amount of the shift and the
renormalization may depend on the distance of the soliton propagation.
The corresponding wavelength $l_c$ is always close to $a$ as discussed above.
Therefore, the maximum velocity of the soliton is approximately given by
$K_y$ as
\begin{equation}
v_c\approx c\sqrt{0.01+0.75\frac{K_y}{J}}.
\label{eq: Maximum}
\end{equation}
\par
%%%%%%%%%%%%%%%%%%%%%%%%%%%%%%%%%%%%%%%%%%%%%%%%%%%%%%%%%%%%%%%%%%%%%%%%%%%%%%%
%
%%%%%%%%%%%%%%%%%%%%%%%%%%%%%%%%%%%%%%%%%%%%%%%%%%%%%%%%%%%%%%%%%%%%%%%%%%%%%%%
It is possible to generate similar solitons when the magnetization at the
edge of the array ($\vec M_0$) is simply rotated and flipped by an external
force.
In fact, the average velocity of the solitons generated in such a way
is plotted in Fig.\ \ref{fig: Velocity} (b) as a function of the flipping time.
When flipping is faster than $M/J\nu$, the average velocity is fixed 
without regard to the flipping time.
The corresponding wavelength of the soliton is also fixed to $a$, and the
velocity is approximately given by Eq.\ (\ref{eq: Maximum}) again.
For slower flipping time ($>\!3M/J\nu$), the average velocity gradually
decreases with the increase of the flipping time, and we have checked that
the soliton can be generated even for much slower flipping time
($\sim\!100M/J\nu$).
\par
%%%%%%%%%%%%%%%%%%%%%%%%%%%%%%%%%%%%%%%%%%%%%%%%%%%%%%%%%%%%%%%%%%%%%%%%%%%%%%%
%
%%%%%%%%%%%%%%%%%%%%%%%%%%%%%%%%%%%%%%%%%%%%%%%%%%%%%%%%%%%%%%%%%%%%%%%%%%%%%%%
Finally, we mention the actual values of parameters for an experimental setup.
It is better to align particles closely ($a\!\sim\!d$) in order to gain
a large $J$ as discussed before.
Since the particle is a single domain, $M\propto d^3$, and thus
$J\propto M^2/a^3 \propto d^3$.
Therefore, $J$ decreases with the decrease of the size of the particle.
For a cylinder of Fe with $d=20$ nm and a height of $10$ nm, however, $J$ is
as large as $J\sim2\times 10^4$ K.
The soliton propagates to the nearest-neighbor site in a time of
$t_0/\sqrt{3K_z/J}$ with $t_0\!\equiv\!M/(J\nu)$, which does not depend
too much on $d$.
For an Fe-particle of the above size, 
$t_0/\sqrt{3K_z/J}\!\sim\!2\times 10^{-11}$ sec ($\sim 50$ GHz)
assuming $K_z/J\!=\!5$.
The maximum velocity in the continuum limit is
$c=\sqrt{3K_z/J}(a/t_0)\sim 1\times 10^3$ m/s for $a=20$ nm.
When the particles are fabricated to have a shape anisotropy of
$K_y\!\sim\!0.7J$, the actual velocity is about 70\% of $c$ according to
Eq.\ (\ref{eq: Maximum}).
Since the clock frequency shown above is proportional to the magnetization,
it might be able to exceed 100 GHz for materials with a large magnetization.
\par
%%%%%%%%%%%%%%%%%%%%%%%%%%%%%%%%%%%%%%%%%%%%%%%%%%%%%%%%%%%%%%%%%%%%%%%%%%%%%%%
%
%%%%%%%%%%%%%%%%%%%%%%%%%%%%%%%%%%%%%%%%%%%%%%%%%%%%%%%%%%%%%%%%%%%%%%%%%%%%%%%
To conclude, we have shown for the first time that a soliton excitation mode
is present in magnetic nano-particle arrays with the dipolar interactions for
both configurations of the magnetization (type I and type II) giving
the minimum total energy depending on the shape anisotropy.
The soliton mode in the type I configuration exists in a wider range of the
shape anisotropy than that in the type II configuration.
The solitons with the maximum velocity, which depend on the shape anisotropy,
always have a wavelength close to the distance between particles.
These solitons are stable independent of the initial conditions, and
they can be generated by fast flipping of the magnetization at the edge of
the array.
The soliton can propagate from a particle to a neighbor particle at a clock
frequency even faster than 100 GHz, and may be feasible in an application
as a signal transmission line.
\par
%%%%%%%%%%%%%%%%%%%%%%%%%%%%%%%%%%%%%%%%%%%%%%%%%%%%%%%%%%%%%%%%%%%%%%%%%%%%%%%
%
%%%%%%%%%%%%%%%%%%%%%%%%%%%%%%%%%%%%%%%%%%%%%%%%%%%%%%%%%%%%%%%%%%%%%%%%%%%%%%%
Finally, it should be noted that the dominant mechanism of the dumping
in the nano-particle arrays and the effects of the disorder in the alignment
of the particles and its shape are not clarified yet.
Further intensive studies will be necessary on these effects.
\par
%%%%%%%%%%%%%%%%%%%%%%%%%%%%%%%%%%%%%%%%%%%%%%%%%%%%%%%%%%%%%%%%%%%%%%%%%%%%%%%
%
%%%%%%%%%%%%%%%%%%%%%%%%%%%%%%%%%%%%%%%%%%%%%%%%%%%%%%%%%%%%%%%%%%%%%%%%%%%%%%%
We would like to thank Takeshi Honda, Jun-ichi Fujita, and Toshio Baba for
helpful discussions.
%%%%%%%%%%%%%%%%%%%%%%%%%%%%%%%%%%%%%%%%%%%%%%%%%%%%%%%%%%%%%%%%%%%%%%%%%%%%%%%
%
%
%
%
%
%%%%%%%%%%%%%%%%%%%%%%%%%%%%%%%%%%%%%%%%%%%%%%%%%%%%%%%%%%%%%%%%%%%%%%%%%%%%%%%

%%%%%%%%%%%%%%%%%%%%%%%%%%%%%%%%%%%%%%%%%%%%%%%%%%%%%%%%%%%%%%%%%%%%%%%%%%%%%%%
%
%
%
%
%
%%%%%%%%%%%%%%%%%%%%%%%%%%%%%%%%%%%%%%%%%%%%%%%%%%%%%%%%%%%%%%%%%%%%%%%%%%%%%%%
\end{document}